# Earth's Mass Variability

Ramy Mawad
Al-Azhar University, Faculty of Science, Astronomy and Meteorology department

Email: ramy@azhar.edu.eg

**Abstract**

The perturbation of the Earth caused by variability of Earth's mass as additional reason with gravity of celestial bodies and shape of the Earth. The Earth eating and collecting matters from space and loss or eject matters to space through its flying in the space around the Sun. The source of the rising in the global sea level is not closed in global warming and icebergs, but the outer space is the additional important source for this rising. The Earth eats waters from space in unknown mechanism. The mass of the Earth become greater in November (before transit apoapsis two months), and become latter in February (after transit apoapsis to two months).

## 1. Introduction

The Earth is the third planet from the Sun. It ranks fifth in size, and its mass is found to be about 5.98 × 1024 kg. Mass is a characteristic that is inherent, and it is independent of the object's environment and the method used to measure it. It is a scalar quantity, that is, a single value with and appropriate unit that has no direction. The mass of the Earth may be determined using Newton's law of gravitation. It is given as the force *F*, which given by $F = G\frac{m_1 \times m_2}{r^2} = m \times a$, where *G* is the Gravitational constant, $m_1$ is the mass of the planet, $m_2$ is the mass of the object, *r* is the radius of the planet, and *a* is the acceleration. Then we can estimate the mass of the Earth as $5.96 \times 10^{24}$ kg.

The Earth's mass have many values depending on the author's estimation method, it is equal to $6.0 \times 10^{24}$ kg (Giancoli, Douglas C., 1980), $6.0 \times 10^{24}$ kg (Larouse, 1981), $5.83 \times 10^{24}$ kg (Smith, Peter J., 1986), $5.98 \times 10^{24}$ kg (Hewitt, 1987), $5.98 \times 10^{24}$ kg (Beichner, Robert J. et. al., 2000), $6.0 \times 10^{24}$ kg (The World Book Encyclopedia. Vol. 6. Chiacago: World Book Inc., 2001), $5.972 \times 10^{24}$ kg (Arnett, Bill, 2002), and $M = a\,r^2/G = 5.98 \times 10^{24}$ kg (Christine Lee, 2002).

The Earth gains mass each day, as a result of incoming debris from space. This occurs in the forms of "falling stars", or meteors, on a dark night. The actual amount of added material depends on each study, though it is estimated that 10 to the 8$^{th}$ power kilograms of in-falling matter accumulates every day. The seemingly large amount, however, is insignificant to the Earth's total mass. The Earth adds an estimated one quadrillionth of one percent to its weight each day (Samantha Dong, 2002).

The flux of solar thermal energy incident on any point at the top of the Earth's atmosphere is a function of the solar constant and of instantaneous values of the Earth's axis of rotation, Each of these parameters changes with time to modulate the latitudinal distribution of insolation (Borisenkov et. al., 1985). Borisenkov et. al. 1985 combined the calculated effects of short and long period orbital perturbations with modeled effects of recorded sunspot and facular activity to examine patterns of terrestrial insolation.

Astronomical variation or variation of orbital parameters can be effect, the orbital parameters of Earth effect on the climate (Sharaf and Budnikova, 1969). In past, many studies assumed that there were no variations in the output of the sun itself; i.e., that solar constant is constant. Recent spaceborne measurements of the total solar irradiance have shown, whoever, that "solar constant" varies from day to day by as much as 0.3% as result of modulation by sunspots and bright faculae (Willson, 1984).

The Earth is move around the sun in ellipse orbit, position of the sun being at a focus of the ellipse. The sun-earth distance depending on sun and earth masses according to newton's law of gravitation, $\mathrm{r}^2 = \frac{\mathrm{G}}{\mathrm{F}}\mathrm{m_s m_e}$, where r the sun-earth distance, G constant of gravitational force of attraction, F the gravitational force of attraction, $m_s$ mass of the sun, and me mass of the earth. The sun-earth distance perturbing from elliptical orbit, theoretically, everybody on the solar system provides non-zero force acting on our earth's orbit; two bodies dominate when computing earth orbit perturbations, it is the main source of the earth's orbital perturbation. The Earth is not a sphere; it is an oblate spheroid. There are also regions of higher mass and higher density called mascons that also introduce irregularities; it is another reason for orbital perturbation.

Those reasons considered theoretically by numerous studies. The author's studies not found that the perturbation of Earth's orbit is closed to forces from everybody and Earth's shape, not empirical or experimental studies proved it. In addition the authors assumed that the Earth's mass is constant and neglected the smallest increasing value of Earth's mass; in addition no author proved that the Earth's mass may be decrease. When Earth's mass exchange it is lead to Earth's orbital perturbation according to newton's law. It is my aim in this paper, I aim is studying the relation between orbital perturbation of the Earth with Earth's mass exchanges, but who can we done it?

The Earth contain land, water, and air, the total mass of the Earth is equal to summation of its components, the changeable of mass of any portion of the earth leads to a changing in the total Earth's mass.

The scientists found changing in the atmosphere, it is reacting with solar wins, unfortunately not accurate measurements or estimation for atmospheric mass and it is so difficult to trace the changeable in the mass or motion of the atmosphere. Then I prefer to neglect the changeable of atmospheric mass it in current study. I can glance that no changeable in the Earth's land and no measurements for it, then we can assume in current study that the mass of the earth is constant. But the water in it different images it is have changeable, many activity and we show it. Fortunately, we have observation for sea level and ices of earth's poles.

Most of this rise can be attributed to the increase in temperature of the sea and the resulting slight thermal expansion of the upper 500m of sea water. Additional contributions, as much as one-fourth of the total, come from water sources on land, such as melting snow and glaciers and extraction of groundwater for irrigation and other agricultural and human needs (Bindoff et. al., 2007). Sea level in last 100 years has been increasing at an average rate of about 1.8 mm per year (Douglas, 1997).

Observational and modeling studies of mass loss from glaciers and ice caps indicate a contribution to sea-level rise of 0.2 to 0.4 mm/yr averaged over the $20^{th}$ century. Over this last million years, whereas it was higher most of the time before then, sea level was lower than today.

Scientists previously had estimated which is greater, ice going in or coming out, called the mass balance, important because it causes changes in global sea level. High-precision gravimetry from satellites in low-noise flight has since determined that in 2006, the Greenland and Antarctic ice sheets experienced a combined mass loss of 475 ± 158 Gt/yr, equivalent to 1.3 ± 0.4 mm/yr sea level rise. Notably, the acceleration in ice sheet loss from 1988-2006 was 21.9 ± 1 $Gt/yr^2$ for Greenland and 14.5 ± 2 $Gt/yr^2$ for Antarctica, for a combined total of 36.3 ± 2 $Gt/yr^2$. This acceleration is 3 times larger than for mountain glaciers and ice caps 12 ± 6 $Gt/yr^2$ (Rignot et. al, 2011).

From previous studies we found that the numerous of studies assumed the main source of sea level rising is polar ice caps and glaciers because the balance of Earth's mass, they assumed that the total Earth's mass is constant. But here in this study I will assumed that the Earth's mass is variable, then the changing in global sea level as a portion of total Earth's mass leads to Earth's mass change as additional reason of changing or rising of sea level like changing in mountain glaciers and ice caps.

## 2. Data Selection

Geocentric Ephemeris data during the period 1995-2006 of the Sun is used in this paper from Astronomical Ephemeris Data from NASA data center (resolution is two days):

http://eclipse.gsfc.nasa.gov/TYPE/ephemeris.html

Sea level data for the TOPEX/Poseidon satellite altimeters based level series (joint mission of NASA and CNES). Satellite data were adjusted to give the same average level as the tide gauges. it downloaded in the same period 1995-2006 from (resolution is ten days):

http://sealevel.colorado.edu/current/sl_ib_ns_global.txt

## 3. Algorithm

We need to compare the perturbations in Earth's orbit (*Distance Shift*) with oscillation in global sea level (*Sea Level Shift*). It is done by following steps.

### 3.1. Estimating the Global Sea Level Shift:

- Calculate the linear fitting of global sea level during the total period from given satellite data. We found it:

$$h = -6445.7 + 3.2305 \times y \quad , R=0.97595 \qquad ...(1)$$

Where *y* is the year including months and days as fraction in the year, and *h* is the sea level in millimeter.

- Completing the date to become daily by using Lagrange's Interpolation Method.

- Subtract the sea levels from linear fitting (equation 1) to get the perturbation only "Global *Sea Level Shift*" as following formula:

$$\Delta\mathrm{h} = \mathrm{H} - \mathrm{h} \qquad ...(2)$$

Where *H* is the observed value of Global sea level and *h* is the calculated value of linear fitting of the seal levels with time as in equation (1).

- List the maximum and minimum peaks in global sea level in given period and record the amplitude $\boldsymbol{A_{sea}}$.

### 3.2. Estimating the Shift of Sun's Distance:

- Read the Sun's distance ***R*** from Geocentric Ephemeris data and complete it to become daily by using Lagrange's Interpolation Method.

- Estimate the distance in elliptical orbit of the Earth around the Sun by using *kepler's* equation as follows:

*T = 367*Y - 7 * (Y + (M+9)/12 ) / 4 + 275*M/9 + D - 730530*

*T = intval(T) + UT/24.0* ...(3)

Where *T* is time scale and *Y* is year, *M* is month and *D* day. The eccentricity of the Earth's orbit *e* (0=circle, 0-1=ellipse, 1=parabola):

$$e = 0.016709 - 1.151\times10^{-9} \times T \qquad ...(4)$$

The mean anomaly of the sun *M* (0 at perihelion; increases uniformly with time):

$M = 356.0470 + 0.9856002585 \times T$ ...(5)

The eccentric anomaly $E$:

$$E = M + e \sin M \ (1 + e \cos M) \quad \text{...(6)}$$

Then the elliptical distance $\boldsymbol{r}$ can be calculated from:

$$r = \sqrt{(\cos E - e)^2 + \left(\sqrt{1 - e^2} \times \sin E\right)^2} \quad \text{...(7)}$$

- Now we ready to get the *distance shift* $\Delta r$ between distance which is recorded in Geocentric Ephemeris data $\boldsymbol{R}$ and elliptical orbital distance $\boldsymbol{r}$ as follows:

$$\Delta r = r - R \quad \text{...(8)}$$

- List the maximum and minimum peaks of *distance shift* $\Delta r$ and record the amplitude $\boldsymbol{A_{dist}}$.

### 3.3. Sea Level Shift-Distance Shift comparisons:

- List the combination the sea level shifts and distance shifts in the same table.

- filter the result data as following cases:

1. Find the sea level peaks which is happens after distance peaks, and which is have the same sign.

2. Find the sea level peaks which is happens after distance peaks, and which is have different sign.

3. Find the distance peaks which is happens after sea level peaks, and which is have the same sign.

4. Find the distance peaks which is happens after sea level peaks, and which is have different sign.

- Plot the time series for distance shift and sea level shift.

- Plot the amplitudes of associated distance shift $\boldsymbol{A_{dist}}$ and sea level shift $\boldsymbol{A_{sea}}$ for all cases.

## 4. Result and conclusions

From figure 1 we show that the sea level is rising with time as in linear fitting which is given in equation 1. When we plot the perturbation of the Earth's orbit with oscillation in the sea level with the time as shown in figure 2.1, we found that the behavior of the both curves are very link in many parts, sometimes are parallels and sometimes are contrary, the figures from 2.2 to 2.4 gives more details in short periods.

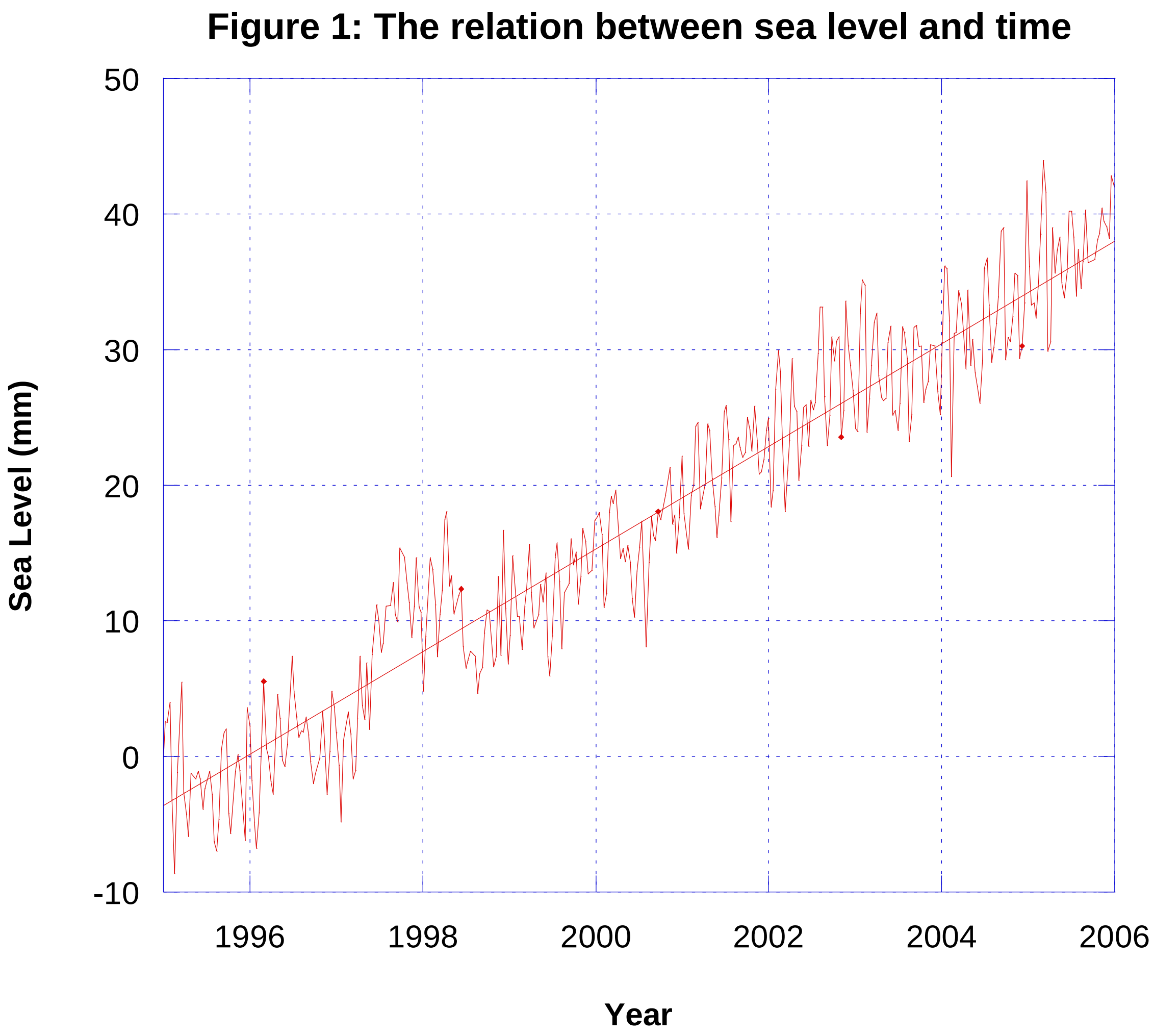

Figure 1: The relation between sea level and time
Sea Level (mm)
50
40
30
20
10
0
-10
1996
1998
2000
2002
2004
2006
Year

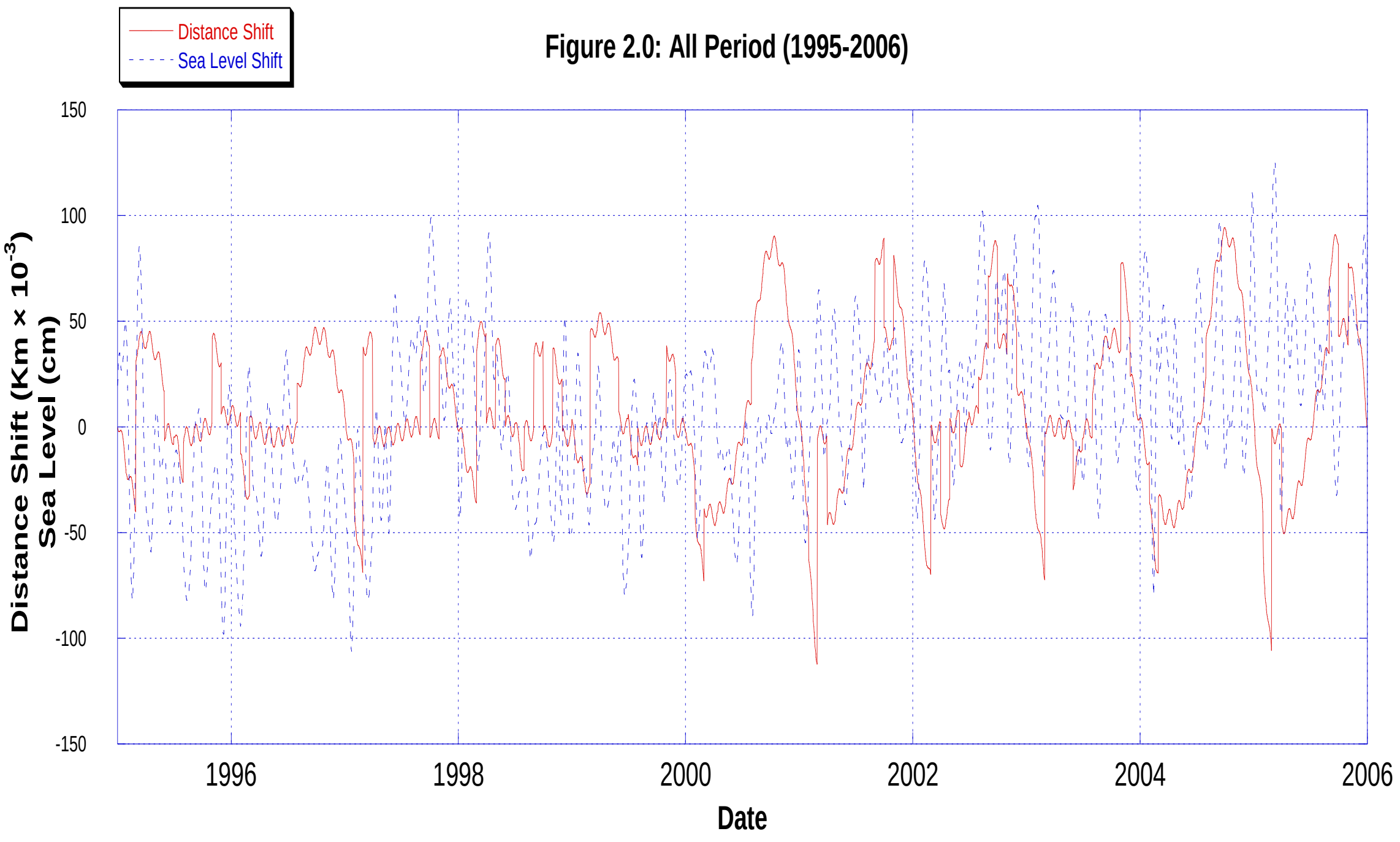

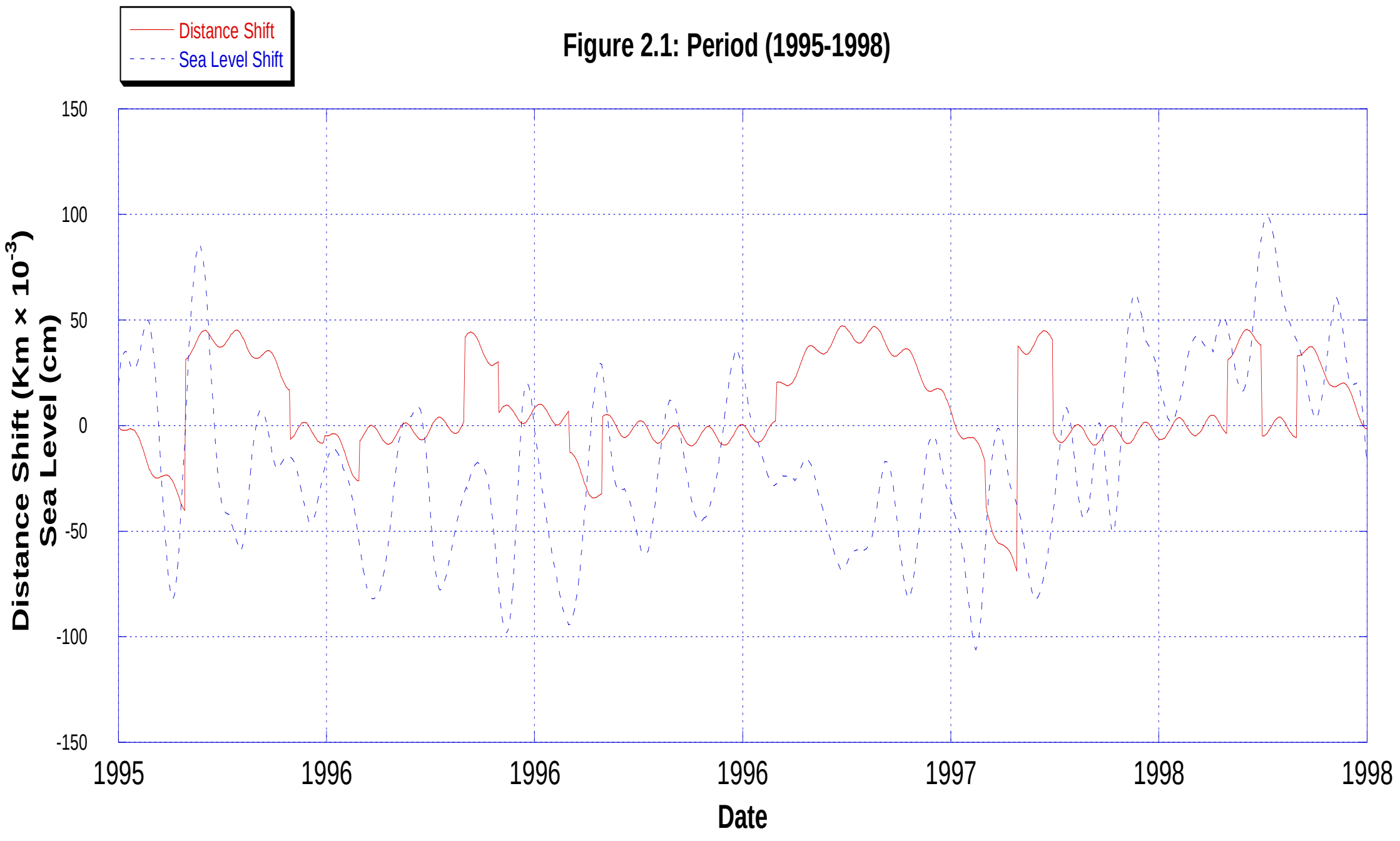

Figure 2.2: Period (1998-2001)

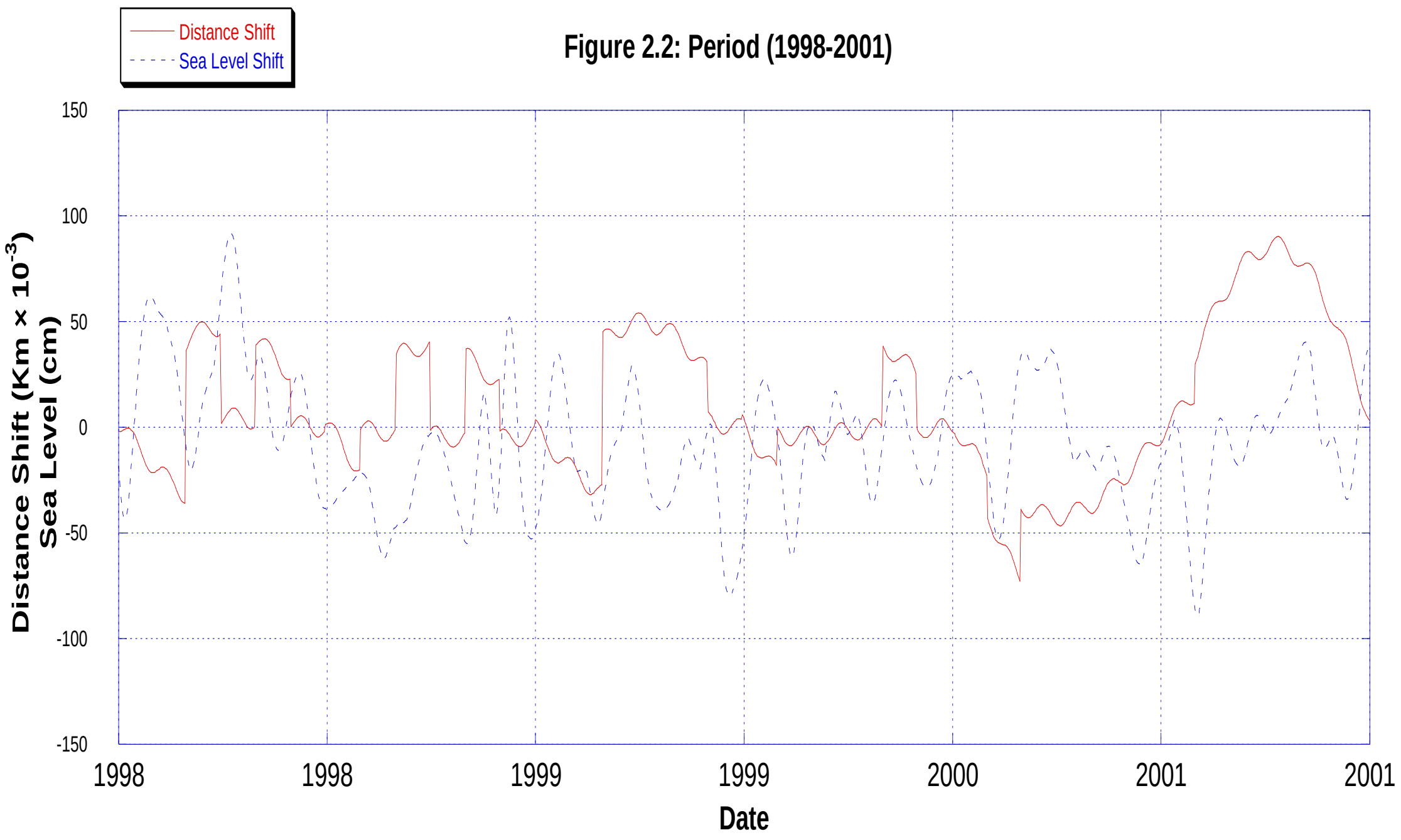

Figure 2.3: Period (2001-2004)

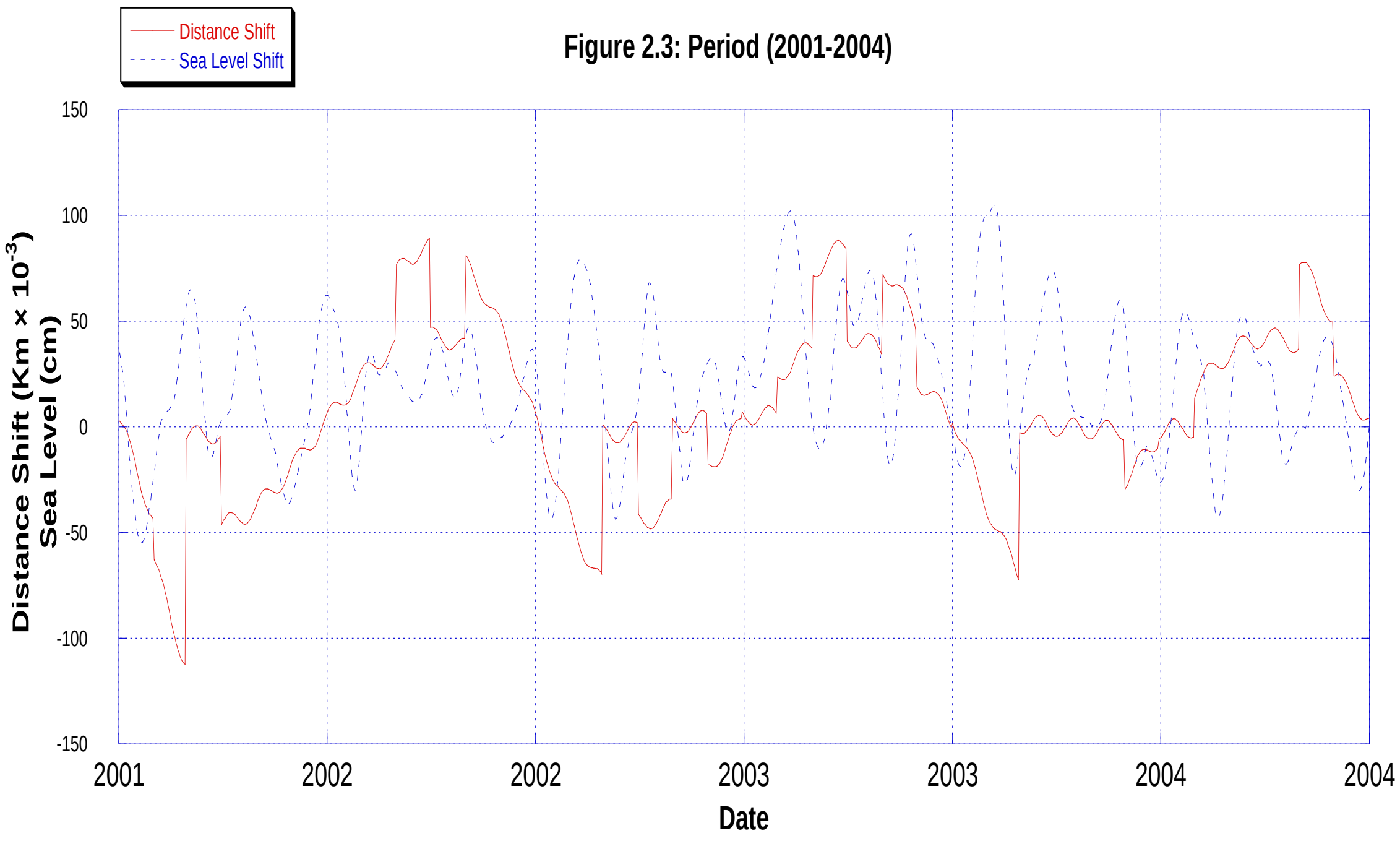

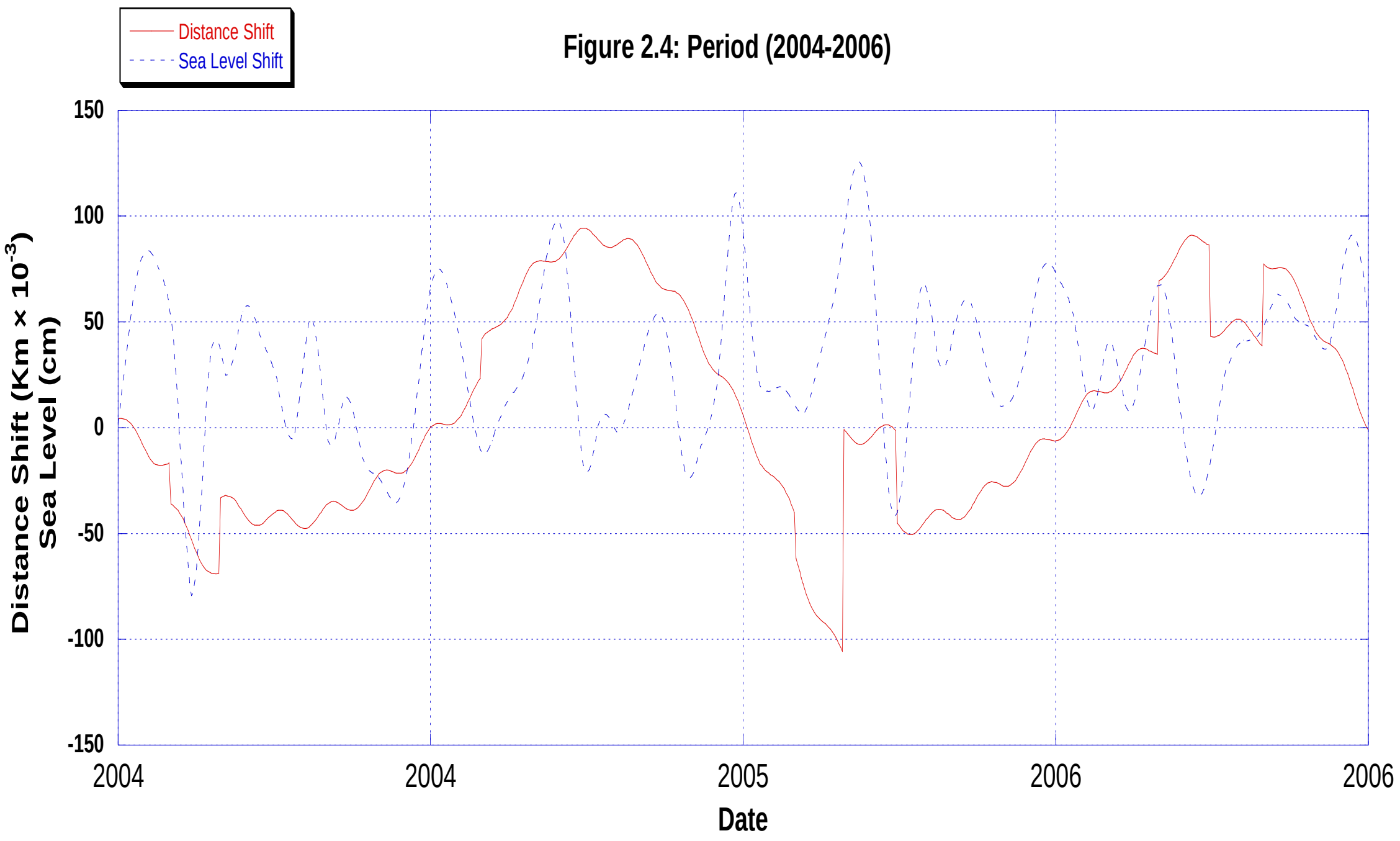

We can show that the amplitude of the perturbation of Earth's orbit increasing with time like rising of the sea level, it is indicate that the perturbation of Earth's orbit is correlated to the sea level. By plotting the relation between amplitudes of the main peaks of the perturbation of the Earth's orbit with fitting value which is given from linear fitting of the sea level which is given from equation 1 at the same time of the amplitude as shown in figure 3, we can found a strong correlation coefficient between them, the result date listed in table 1.

$$H = -6.124 + 0.35655\ A,\ R=0.77521 \qquad ...(9)$$

Where $H$ is the fitting value of the sea level, and H is the $A$ is the amplitude of the main peaks of the perturbation of the Earth's orbit.

Mostly, we can notice that the main peaks happens two times per the year, the first peak is through and happen in around February month, and second peak is crest happens in November.

Now, we want to study the all peaks of the orbital perturbation with the sea level.

We want to study the peaks of sea level shifts and distance shifts in the four cases as shown in figures from 4.1 to 4.4, I found 297 peaks of sea level shits happens in the selected period and 254 peaks of distance shifts happens in the selected period, we found the good correlations between sea level and perturbation in Earth's orbit and oscillation in sea level as follows:

1. In the case of sea level peak happens after distance peak and which is have the same sign: the correlation is $R=0.577$ of $\boldsymbol{A_{sea}}=9.5699 + 4.303\ \boldsymbol{A_{dist}}$ for 94 events from 297 events (32%) (figure 4.1).

2. In the case of sea level peak happens after distance peak and which is have different sign: the correlation is $R=0.54748$ of $\boldsymbol{A_{sea}}=0.4371 + 0.064354\ \boldsymbol{A_{dist}}$ for 96 events from 297 events (32%) (figure 4.2).

3. In the case of distance peaks happens after sea level peak, and which is have the same sign: the correlation is $R=0.58206$ of $\boldsymbol{A_{sea}}=0.47714 - 0.068226\ \boldsymbol{A_{dist}}$ for 91 events from 254 events (36%) (figure 4.3).

4. in the case distance peaks happens after sea level peak and which is have different sign: the correlation is $R=0.5$ of $\boldsymbol{A_{sea}} = 0.38723 + 0.061758\ \boldsymbol{A_{dist}}$ for 94 events from 254 events (37%) (Figure 4.4).

**Figue 3: The relationship between Amplitude of the orbital perturbation of the Earth and fit value of the Sea Level**

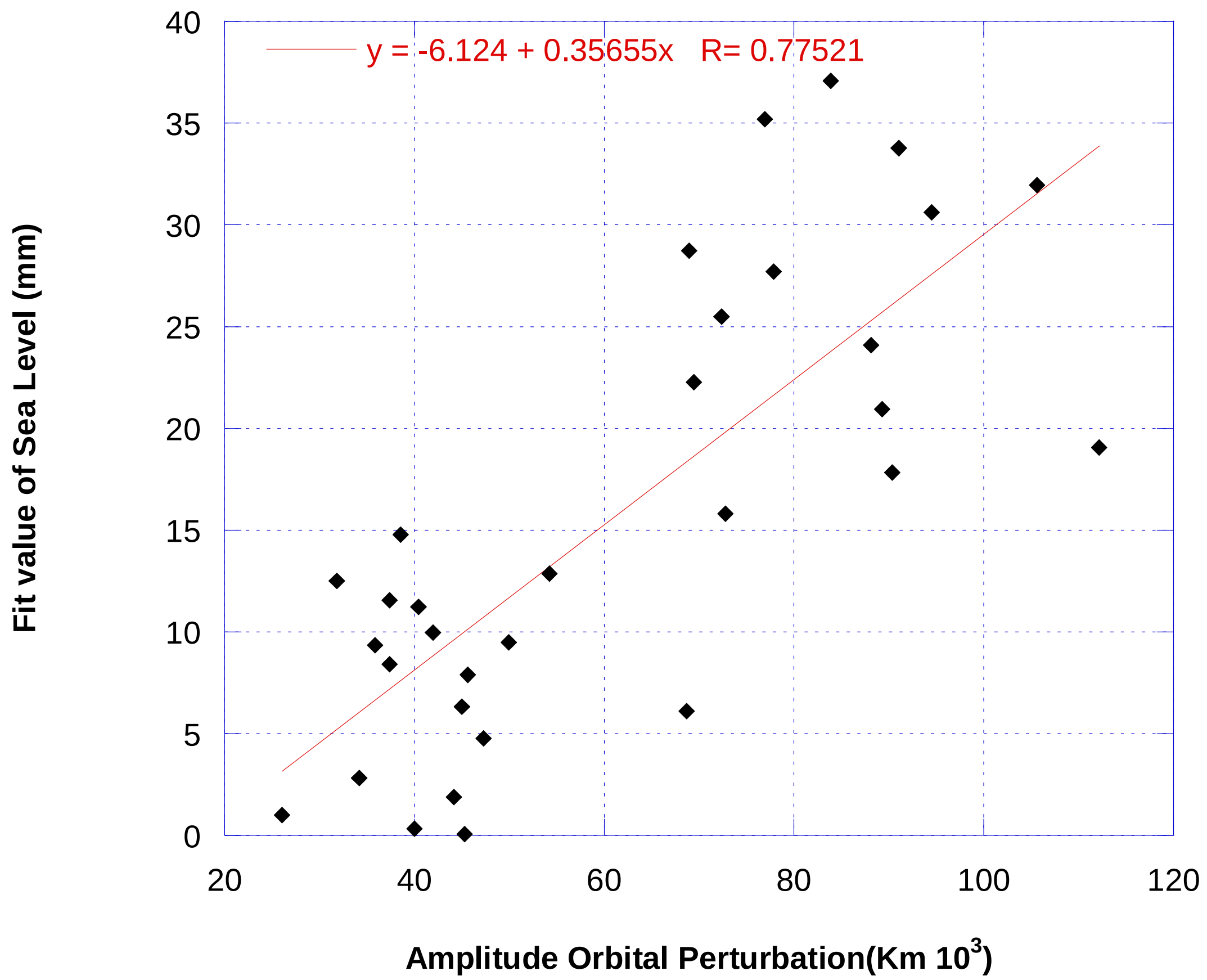

**Figure 4.1: sea level peak happens after distance peak and which is have the same sign**

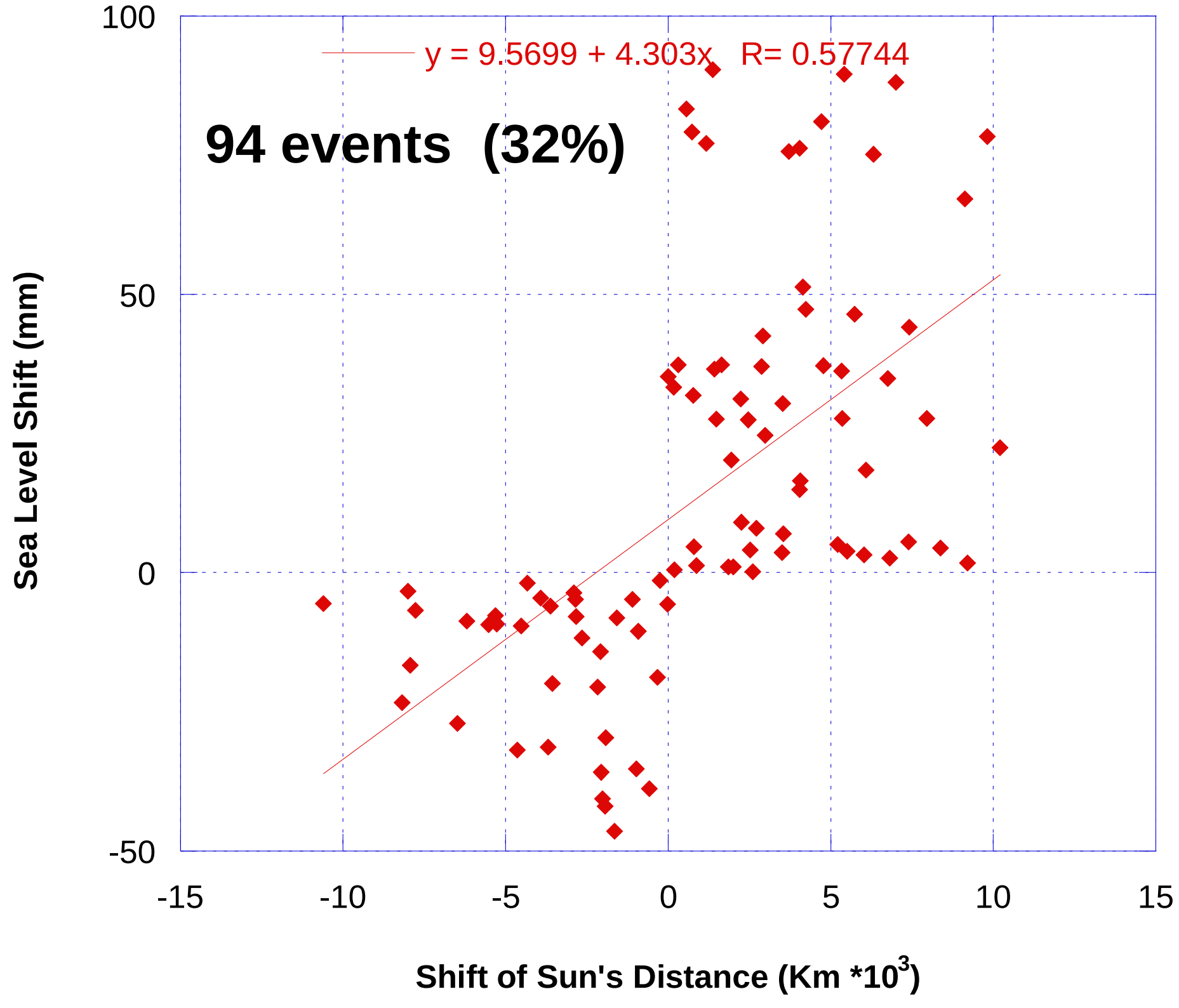

**Figure 4.2: sea level peak happens after Distance peak and which is have different sign**

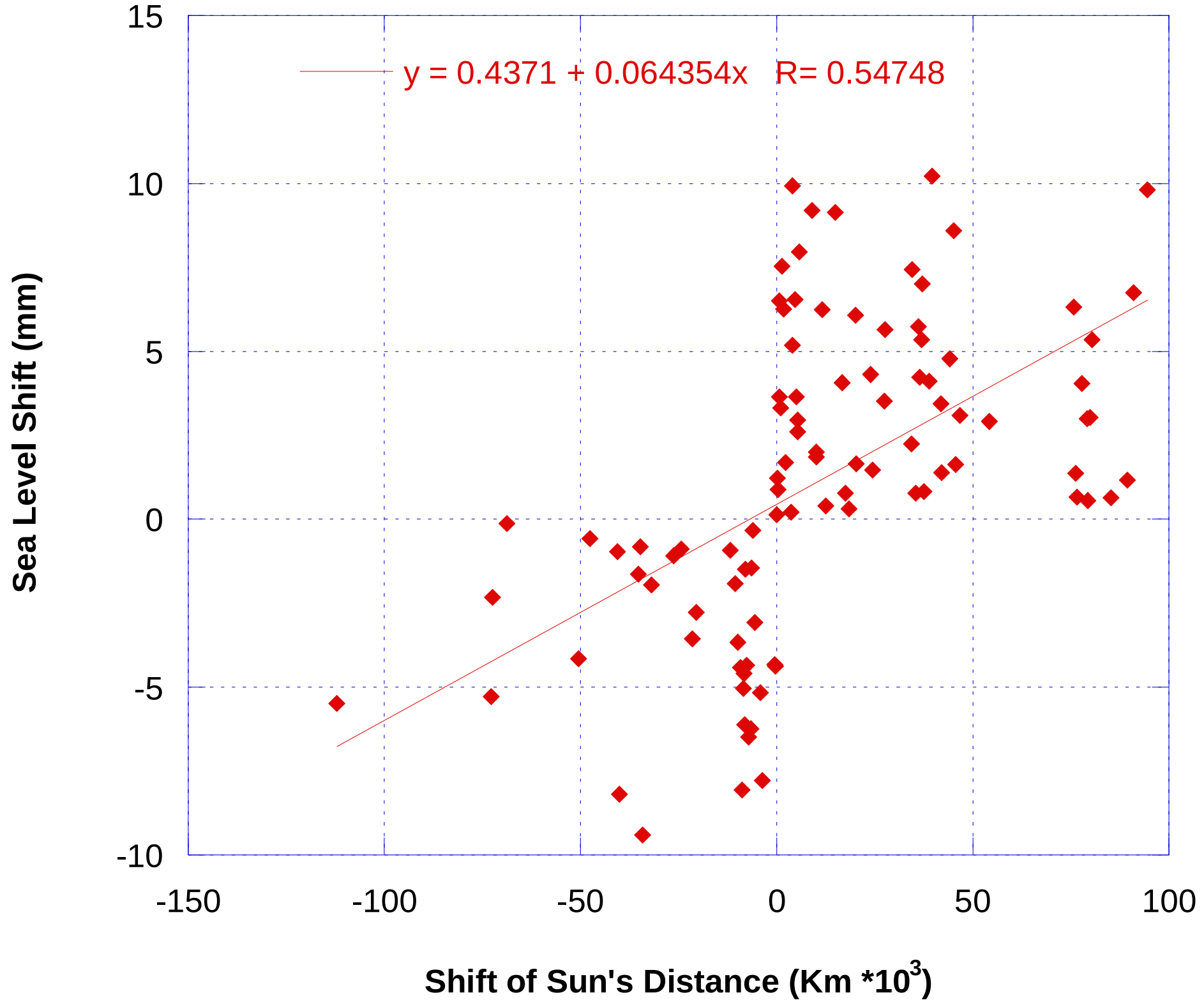

**Figure 4.3: distance peaks happens after sea level peak, and which is have the same sign**

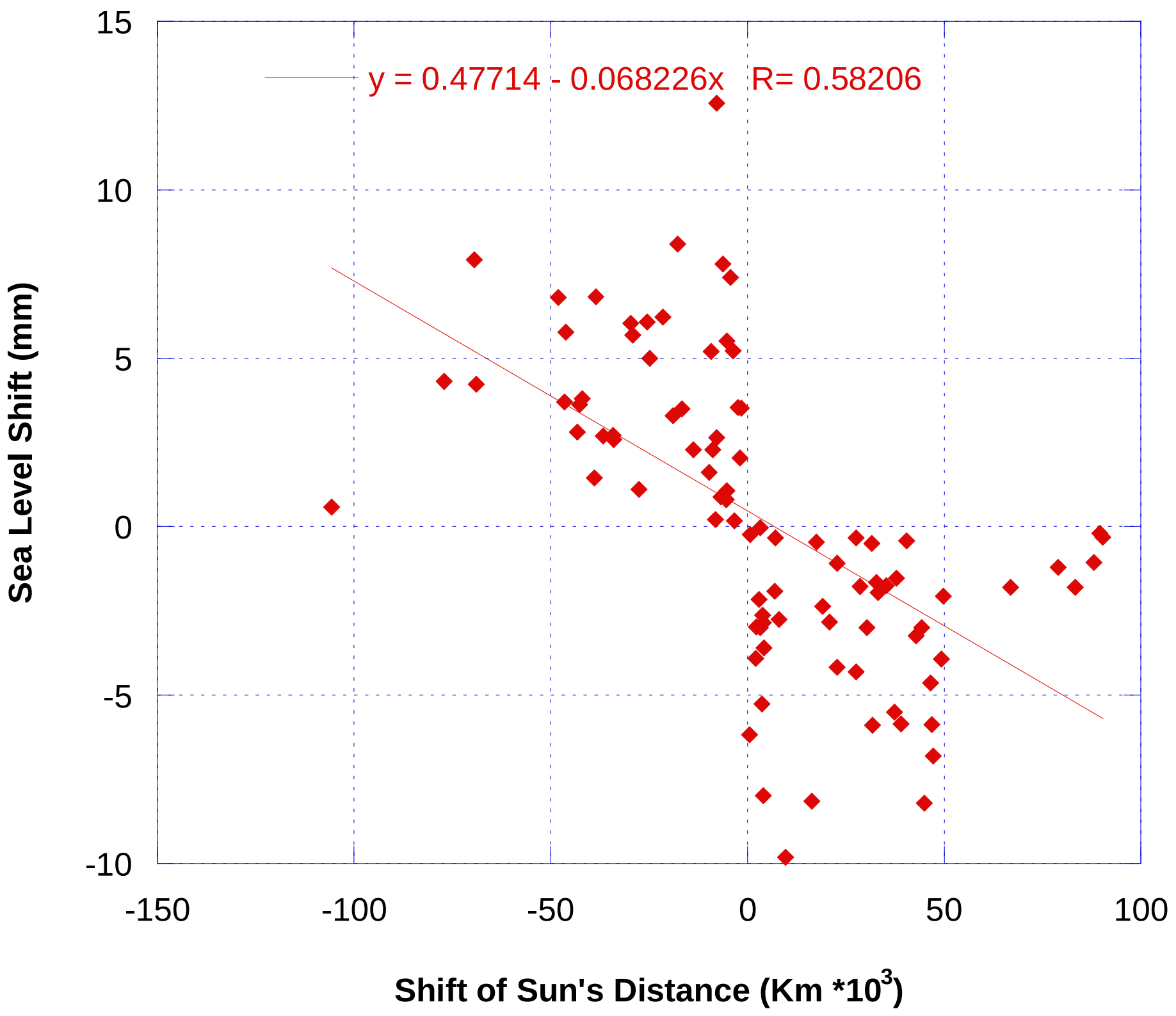

**Figure 4.4: distance peaks happens after sea level peak and which is have different sign**

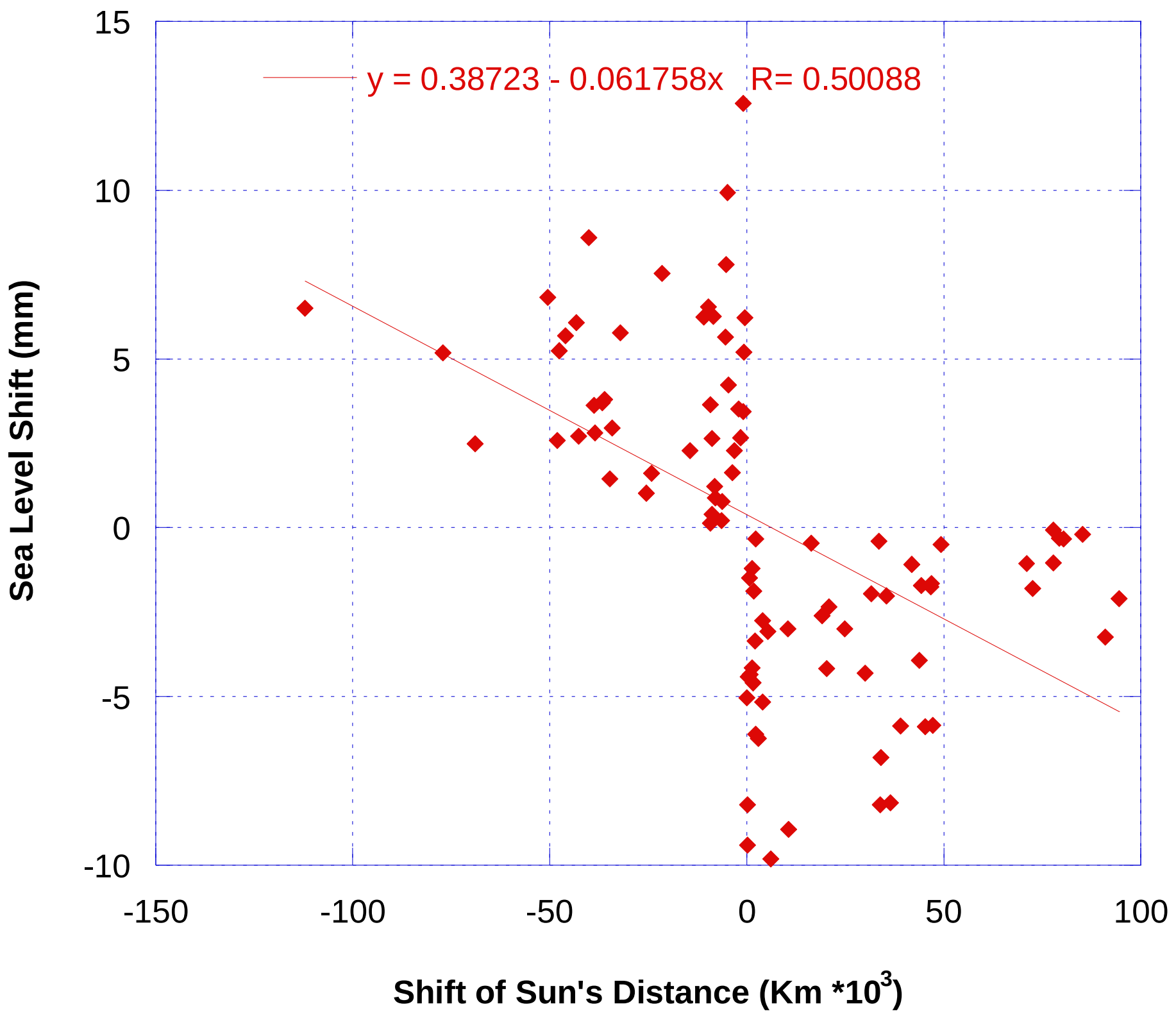

## 5. Conclusions

The oscillation in the global sea level have strong correlation with perturbation of the Earth's orbit (figures 2.x, 3, 4.x), then the oscillation in global sea level indicating to the Earth's mass is variable according to Newton's law of gravitation as additional reason of sea level rising.

The perturbation of the Earth caused by variability in Earth's mass as additional reason with gravity of celestial bodies and shape of the Earth.

The Earth eating and collecting matters from space and loss or eject matters to space through its flying in the space around the Sun.

The source of the rising in the global sea level is not closed in global warming and icebergs according to mass level, but the outer space is one of sources for this rising and may be it is the main source.

If the oscillation of sea level indicating to variability in Earth's mass then the Earth eating waters from space in unknown mechanism, then we can say if the peaks of global sea level and perturbation in Earth's orbit are associated, then this indicate to the mass of the Earth is varying, then the Earth collect or loss water from or to space. When the peaks are not associated then the other peaks caused by other reasons.

We have two types for peaks of Earth's mass variability (perturbation):

***1. Main peaks:*** The smallest velocity of the Earth in apoapsis, after this point the Earth moving toward the Sun in its orbit with increasing in velocity and opposite direction of solar wind, in this case the Earth eating and collecting matters from space greater than losing matters, then the mass is increasing in this epoch, when Earth reaching to November (before apoapsis) the mass of the Earth become greater (crest peak). After November (transit apoapsis) the Earth's velocity and Sun's gravity become to greater values, the Sun's gravity with Earth's velocity causing suddenly motion for the Earth, it is leads to losing mass from Earth into space, the earth eject matters quickly, the mass of the Earth become smallest value in February (through peak). Mostly it is happen two times in beginning and last of the year and the oscillation curves moving parallel.

***2. Small peaks:*** It is happens in remainder parts of the year, we have two cases: 1) *Parallel case*: the both curves moving parallels, the crest peaks of the both curves happen together and the through peaks happen together. 2) *Opposite case*: the both curves moving opposite, crest peak of the sea level shift happens with through peak of distance shift and it is follows through peak of the sea level shift happens with crest peak of distance shift.

The type of *small peaks* need more studying, why some time parallel, and why some time opposite?

The reasons: 1) data of the global sea level and Sun's distance is not daily, I depended on interpolations to predict the unavailable value, 2) in addition I used the daily maximum values as a peaks, it is not accrued method, 3) estimating the eccentric anomaly simple formula without iteration, 4) and not all of the peaks of the global of sea level caused closely by variability in the Earth's mass but it causing by additional reasons too at the same time. Those reasons affecting in correlation between global sea level and orbital perturbation.

## 6. Acknowledgment

The authors thank the NASA data center and TOPEX/Poseidon for providing us the data.

**Table 1:** List of main peaks of the amplitude in Earth's orbit with fitting value of the sea level.

| Amplitude | SeaLevel | Date |
|---|---|---|
| 40.0292442444 | 0.339160273972 | 1995-02-28 |
| 45.2755881241 | 0.0679712328774 | 1995-04-15 |
| 26.0796514021 | 1.01499452055 | 1995-07-31 |
| 44.2110173559 | 1.88236164384 | 1995-11-06 |
| 34.1766147514 | 2.83697814208 | 1996-02-22 |
| 47.2822037801 | 4.76998224044 | 1996-09-28 |
| 68.7268830195 | 6.12183972603 | 1997-02-28 |
| 45.0336514618 | 6.33425616438 | 1997-03-24 |
| 45.6352132941 | 7.90967808219 | 1997-09-18 |
| 37.3720406656 | 8.40531643836 | 1997-11-13 |
| 35.8564374612 | 9.35233972603 | 1998-02-28 |
| 49.9340052634 | 9.4851 1998.2 | 1998-03-15 |
| 41.9530823067 | 9.97188767123 | 1998-05-09 |
| 40.4519807461 | 11.2463863014 | 1998-09-30 |
| 37.3635240676 | 11.547309589 | 1998-11-03 |
| 31.8384980288 | 12.4943328767 | 1999-02-18 |
| 54.2395749145 | 12.8660616438 | 1999-04-01 |
| 38.5562519165 | 14.7601082192 | 1999-11-01 |
| 72.7744707309 | 15.8207636612 | 2000-02-29 |
| 90.3771521786 | 17.8243797814 | 2000-10-13 |
| 112.156055434 | 19.043839726 | 2001-02-28 |
| 89.3350029705 | 20.9378863014 | 2001-09-30 |
| 69.4748318445 | 22.274339726 | 2002-02-28 |
| 88.1238754539 | 24.1064315068 | 2002-09-23 |
| 72.3746106634 | 25.504839726 | 2003-02-28 |
| 77.8877276214 | 27.7175109589 | 2003-11-05 |
| 68.9716840449 | 28.7339371585 | 2004-02-28 |
| 94.5541886864 | 30.6316352459 | 2004-09-30 |
| 105.639822604 | 31.965839726 | 2005-02-28 |
| 91.0594360335 | 33.7713794521 | 2005-09-20 |
| 76.9812688836 | 35.196339726 | 2006-02-28 |
| 83.8822270225 | 37.0903863014 | 2006-09-30 |

**Table 1:** Result list of peaks of distance amplitudes and sea level amplitudes

| Date | Distance Amplitude | Sea Level Amplitude |
|---|---|---|
| 1/6/1995 | -2.1048 | |
| 1/7/1995 | | 3.5161 |
| 1/12/1995 | -1.4944 | |
| 1/14/1995 | | 2.6697 |
| 1/27/1995 | | 5.0067 |
| 2/4/1995 | -24.703 | |
| 2/12/1995 | -23.343 | |
| 2/18/1995 | | -8.1917 |
| 2/28/1995 | -40.029 | |
| 3/13/1995 | | 8.5923 |
| 3/18/1995 | 45.144 | |
| 3/31/1995 | 37.19 | |
| 4/15/1995 | 45.276 | |
| 4/19/1995 | | -5.9038 |
| 5/2/1995 | 31.8 | |
| 5/6/1995 | | 0.76713 |
| 5/13/1995 | 35.51 | |
| 5/19/1995 | | -2.0262 |
| 5/28/1995 | | -1.4205 |
| 5/29/1995 | | -1.4709 |
| 5/30/1995 | | -1.4513 |
| 6/1/1995 | -6.412 | |
| 6/13/1995 | 1.6769 | |
| 6/18/1995 | | -4.5954 |
| 6/28/1995 | -8.2123 | |
| 7/1/1995 | -4.66 | |
| 7/2/1995 | -4.8219 | |
| 7/9/1995 | -3.7042 | -1.0993 |
| 7/31/1995 | -26.08 | |
| 8/11/1995 | 0.18032 | |
| 8/12/1995 | | -8.2015 |
| 8/15/1995 | | -8.0708 |
| 8/16/1995 | | -8.0722 |
| 8/26/1995 | -8.7847 | |
| 9/10/1995 | 1.3296 | |
| 9/21/1995 | | 0.88378 |
| 9/24/1995 | -6.7144 | |
| 10/10/1995 | 4.0198 | -7.7767 |
| 10/23/1995 | -3.5656 | |
| 11/2/1995 | | -2.8991 |
| 11/3/1995 | | -2.9861 |
| 11/6/1995 | 44.211 | |
| 11/12/1995 | | -1.7256 |
| 11/13/1995 | | -1.7868 |
| 11/14/1995 | | -1.7791 |
| 11/25/1995 | 28.611 | |
| 11/30/1995 | 30.401 | |
| 12/1/1995 | 6.166 | |
| 12/7/1995 | | -9.8043 |
| 12/8/1995 | 9.746 | |
| 12/22/1995 | 1.0391 | |
| 12/25/1995 | | 2 |
| 1/6/1996 | 10.149 | |
| 1/21/1996 | 0.29162 | |
| 1/31/1996 | 6.9661 | -9.4153 |
| 2/22/1996 | -34.177 | |
| 2/28/1996 | | 2.9541 |
| 3/5/1996 | 5.3738 | |
| 3/16/1996 | | -3.0664 |
| 3/20/1996 | -5.5074 | -2.9794 |
| 4/3/1996 | 2.2888 | |
| 4/7/1996 | | -6.1197 |
| 4/19/1996 | -8.1726 | |
| 4/29/1996 | | 1.2205 |
| 5/3/1996 | 0.15629 | |
| 5/18/1996 | -9.5058 | |
| 5/26/1996 | | -4.5228 |
| 5/29/1996 | | -4.3646 |
| 5/30/1996 | | -4.3675 |
| 6/2/1996 | -0.28073 | |
| 6/16/1996 | -9.2432 | |
| 6/27/1996 | | 3.634 |
| 7/1/1996 | 0.64533 | |
| 7/15/1996 | -7.8556 | |
| 7/29/1996 | | -2.8278 |
| 8/3/1996 | 20.807 | |
| 8/6/1996 | | -2.3469 |
| 8/7/1996 | | -2.3916 |
| 8/8/1996 | | -2.3741 |
| 8/11/1996 | 19.084 | |
| 8/17/1996 | | -2.6054 |
| 8/26/1996 | | -1.5366 |
| 8/31/1996 | 37.936 | |
| 9/11/1996 | 33.986 | |
| 9/27/1996 | | -6.8007 |
| 9/28/1996 | 47.282 | |
| 10/12/1996 | | -5.8667 |
| 10/13/1996 | 39.179 | |
| 10/16/1996 | | -5.8881 |
| 10/26/1996 | 46.958 | |
| 11/4/1996 | | -1.6691 |
| 11/13/1996 | 32.846 | |
| 11/23/1996 | 36.529 | |
| 11/25/1996 | | -8.1504 |
| 12/14/1996 | 16.338 | |
| 12/16/1996 | | -0.45564 |
| 12/21/1996 | 17.623 | |
| 1/13/1997 | -6.2222 | |
| 1/19/1997 | -5.4735 | |
| 1/23/1997 | | -10.61 |
| 2/12/1997 | | -0.13748 |
| 2/28/1997 | -68.727 | |
| 3/1/1997 | 37.732 | |
| 3/9/1997 | 33.938 | |
| 3/17/1997 | | -8.2 |
| 3/24/1997 | 45.034 | |
| 4/8/1997 | -7.9486 | |
| 4/12/1997 | | 0.87588 |
| 4/22/1997 | 0.41525 | |
| 4/28/1997 | | -4.4109 |
| 5/7/1997 | -9.1701 | |
| 5/11/1997 | | 0.13181 |
| 5/22/1997 | 0.11276 | |
| 5/23/1997 | | -5.0459 |
| 6/5/1997 | -8.4548 | |
| 6/12/1997 | | 6.2552 |
| 6/21/1997 | 1.7155 | |
| 7/4/1997 | -6.4744 | |
| 7/13/1997 | | 0.20975 |
| 7/20/1997 | 3.7547 | |
| 8/3/1997 | -4.6763 | |
| 8/5/1997 | | 4.2299 |
| 8/16/1997 | | 3.5792 |
| 8/18/1997 | | 3.6326 |
| 8/19/1997 | 4.9944 | 3.4938 |
| 8/28/1997 | | 5.2192 |
| 8/31/1997 | -3.6805 | |
| 9/14/1997 | | 1.6414 |
| 9/18/1997 | 45.635 | |
| 10/2/1997 | -4.7609 | |
| 10/6/1997 | | 9.9314 |
| 10/16/1997 | 3.9959 | |
| 10/31/1997 | -5.558 | |
| 11/1/1997 | 33.324 | |
| 11/2/1997 | 33.173 | |
| 11/13/1997 | 37.372 | |
| 11/17/1997 | | 0.31919 |
| 12/3/1997 | 18.451 | |
| 12/5/1997 | | 6.0801 |
| 12/11/1997 | 20.255 | |
| 12/20/1997 | | 1.9446 |
| 12/24/1997 | | 2.0286 |
| 1/2/1998 | -1.8497 | |
| 1/6/1998 | | -4.3396 |
| 1/9/1998 | -0.43143 | |
| 1/29/1998 | | 6.2093 |
| 1/31/1998 | -21.55 | |
| 2/9/1998 | -18.825 | |
| 2/28/1998 | -35.856 | |
| 3/5/1998 | | -2.0623 |
| 3/15/1998 | 49.934 | |
| 3/28/1998 | 42.971 | |
| 3/31/1998 | 44.09 | |
| 4/1/1998 | 1.816 | |
| 4/9/1998 | | 9.2056 |
| 4/12/1998 | 9.077 | |
| 4/27/1998 | -0.79241 | 2.2579 |
| 5/3/1998 | | 3.4349 |
| 5/9/1998 | 41.953 | |
| 5/21/1998 | | -1.097 |
| 5/29/1998 | 22.763 | |
| 5/31/1998 | 22.869 | |
| 6/1/1998 | 0.24147 | |
| 6/8/1998 | | 2.609 |
| 6/10/1998 | 5.455 | |
| 6/25/1998 | -4.5505 | |
| 7/2/1998 | | -3.9108 |
| 7/5/1998 | 2.0734 | |
| 7/11/1998 | | -3.3699 |
| 7/12/1998 | | -3.4114 |
| 7/21/1998 | | -2.7525 |
| 7/22/1998 | | -2.7782 |
| 7/27/1998 | -20.504 | |
| 8/1/1998 | | -2.1575 |
| 8/8/1998 | 3.0455 | |
| 8/21/1998 | | -6.2491 |
| 8/22/1998 | -6.5317 | |
| 9/8/1998 | 39.74 | |
| 9/20/1998 | 33.547 | |
| 9/28/1998 | | -0.3933 |
| 9/29/1998 | | -0.42603 |
| 9/30/1998 | 40.452 | |
| 10/1/1998 | -1.4273 | |
| 10/3/1998 | | -0.23984 |
| 10/6/1998 | 0.70933 | |
| 10/21/1998 | -9.3439 | |
| 11/2/1998 | | -5.5082 |
| 11/3/1998 | 37.364 | |
| 11/17/1998 | | 1.6554 |
| 11/22/1998 | 20.308 | |
| 11/27/1998 | | -4.1711 |
| 11/30/1998 | 22.748 | |
| 12/1/1998 | -1.7489 | |
| 12/4/1998 | -0.74036 | |
| 12/9/1998 | | 5.2138 |
| 12/19/1998 | -9.1758 | |
| 12/28/1998 | | -5.2725 |
| 1/1/1999 | 3.6411 | |
| 1/20/1999 | | 3.5021 |
| 1/21/1999 | -16.714 | |
| 1/29/1999 | -14.195 | |
| 2/6/1999 | | -2.0823 |
| 2/12/1999 | | -1.965 |
| 2/18/1999 | -31.838 | |
| 2/25/1999 | | -4.6324 |
| 3/5/1999 | 46.635 | |
| 3/17/1999 | 42.551 | |
| 3/26/1999 | | 2.9122 |
| 4/1/1999 | 54.24 | |
| 4/17/1999 | 43.801 | |
| 4/22/1999 | | -3.9303 |
| 4/29/1999 | 49.245 | |
| 5/14/1999 | | -0.49016 |
| 5/18/1999 | 31.588 | |
| 5/25/1999 | | -1.952 |
| 5/26/1999 | 33.239 | |
| 6/3/1999 | | 0.17439 |
| 6/14/1999 | -3.3088 | |
| 6/20/1999 | | -7.9994 |
| 6/28/1999 | 4.0685 | |
| 6/30/1999 | 3.7877 | |
| 7/1/1999 | 6.1154 | |
| 7/18/1999 | -14.305 | |
| 7/20/1999 | | 2.2789 |
| 7/24/1999 | -13.666 | |
| 7/31/1999 | -17.924 | |
| 8/1/1999 | -0.47325 | |
| 8/12/1999 | -8.7651 | |
| 8/13/1999 | | -6.1905 |
| 8/27/1999 | 0.55116 | |
| 8/30/1999 | | 0.19821 |
| 9/10/1999 | -8.1166 | |
| 9/11/1999 | | -1.5717 |
| 9/21/1999 | | 1.6888 |
| 9/26/1999 | 2.3084 | |
| 10/1/1999 | | -0.34106 |
| 10/10/1999 | -5.9809 | 0.63311 |
| 10/23/1999 | | -3.6149 |
| 10/25/1999 | 4.1199 | |
| 10/31/1999 | 0.46296 | |
| 11/1/1999 | 38.556 | |
| 11/10/1999 | 31.252 | |
| 11/12/1999 | | 2.2393 |
| 11/21/1999 | 34.413 | |
| 12/8/1999 | -4.8277 | |
| 12/9/1999 | | -2.852 |
| 12/23/1999 | 4.0612 | |
| 1/3/2000 | -3.0909 | 2.519 |
| 1/8/2000 | | 2.2919 |
| 1/11/2000 | -8.7307 | |
| 1/17/2000 | | 2.6536 |
| 1/18/2000 | -7.7076 | |
| 2/10/2000 | | -5.2985 |
| 2/29/2000 | -72.774 | |
| 3/1/2000 | -38.627 | |
| 3/4/2000 | | 3.6126 |
| 3/7/2000 | -42.7 | |
| 3/16/2000 | | 2.7129 |
| 3/17/2000 | | 2.7341 |
| 3/18/2000 | | 2.6875 |
| 3/20/2000 | -36.536 | |
| 3/27/2000 | | 3.706 |
| 4/5/2000 | -46.5 | |
| 4/16/2000 | | -1.6432 |
| 4/20/2000 | -35.268 | |
| 4/25/2000 | | -0.96327 |
| 5/2/2000 | -40.603 | |
| 5/6/2000 | | -2.0253 |
| 5/15/2000 | | -0.90759 |
| 5/16/2000 | | -0.91619 |
| 5/17/2000 | | -0.88679 |
| 5/21/2000 | -24.343 | |
| 5/31/2000 | -27.067 | |
| 6/13/2000 | | -6.4726 |
| 6/14/2000 | | -6.4414 |
| 6/15/2000 | | -6.4882 |
| 6/21/2000 | -7.0874 | |
| 6/29/2000 | -8.7613 | |
| 7/14/2000 | | 0.39666 |
| 7/20/2000 | 12.507 | |
| 7/28/2000 | 10.64 | |
| 8/3/2000 | | -8.9379 |
| 8/23/2000 | | 0.44809 |
| 9/8/2000 | | -1.8046 |
| 9/16/2000 | 83.332 | |
| 9/24/2000 | | 0.56288 |
| 9/26/2000 | 79.478 | |
| 10/4/2000 | | -0.31313 |
| 10/13/2000 | 90.377 | |
| 10/21/2000 | | 1.3759 |
| 10/22/2000 | | 1.3575 |
| 10/30/2000 | 76.273 | |
| 11/6/2000 | | 4.0408 |
| 11/7/2000 | 77.724 | |
| 11/22/2000 | | -1.0413 |
| 11/30/2000 | | -0.354 |
| 12/12/2000 | | -3.4108 |
| 12/30/2000 | | 3.6586 |
| 1/21/2001 | | -5.4906 |
| 2/28/2001 | -112.16 | |
| 3/5/2001 | | 6.4953 |
| 3/10/2001 | 0.65129 | |
| 3/23/2001 | | -1.4862 |
| 3/24/2001 | -7.9714 | |
| 3/31/2001 | -4.2144 | |
| 4/1/2001 | -46.161 | |
| 4/9/2001 | -40.441 | |
| 4/21/2001 | -45.925 | |
| 4/22/2001 | | 5.7017 |
| 5/10/2001 | -29.168 | |
| 5/19/2001 | -31.274 | |
| 5/29/2001 | | -3.6745 |
| 6/11/2001 | -9.8353 | |
| 6/18/2001 | -10.836 | |
| 7/2/2001 | | 6.2317 |
| 7/10/2001 | 11.695 | |
| 7/17/2001 | 10.366 | |
| 7/27/2001 | | -2.9895 |
| 8/7/2001 | 30.426 | |
| 8/10/2001 | | 3.5183 |
| 8/17/2001 | 27.421 | |
| 8/18/2001 | | 2.4626 |
| 8/25/2001 | | 3.0262 |
| 9/7/2001 | 79.81 | |
| 9/15/2001 | 77.097 | |
| 9/17/2001 | | 1.1646 |
| 9/30/2001 | 89.335 | |
| 10/1/2001 | 47.117 | |
| 10/2/2001 | 47.352 | |
| 10/6/2001 | | 4.2263 |
| 10/18/2001 | 36.55 | |
| 10/22/2001 | | 1.4147 |
| 10/23/2001 | | 1.4326 |
| 10/24/2001 | | 1.3828 |
| 10/30/2001 | 42.06 | |
| 10/31/2001 | 42.034 | |
| 11/1/2001 | 81.066 | |
| 11/4/2001 | | 4.7056 |
| 11/26/2001 | | -0.73814 |
| 12/2/2001 | | -0.45976 |
| 12/3/2001 | | -0.47634 |
| 12/29/2001 | | 3.6695 |
| 1/15/2002 | | -4.3888 |
| 2/9/2002 | | 7.9293 |
| 2/28/2002 | -69.475 | |
| 3/1/2002 | 0.9228 | |
| 3/12/2002 | | -4.3568 |
| 3/14/2002 | -7.5131 | |
| 3/29/2002 | 2.5926 | |
| 4/10/2002 | | 6.8136 |
| 4/12/2002 | -48.168 | |
| 4/24/2002 | | 2.5859 |
| 4/29/2002 | -34.037 | 2.6994 |
| 4/30/2002 | -34.157 | |
| 5/1/2002 | 3.939 | |
| 5/12/2002 | -2.7574 | -2.7497 |
| 5/27/2002 | 7.9675 | |
| 5/29/2002 | | 2.7098 |
| 5/30/2002 | | 2.6932 |
| 6/4/2002 | | 3.2825 |
| 6/7/2002 | -18.868 | |
| 6/19/2002 | | -0.33224 |
| 7/1/2002 | 7.1183 | 3.3258 |
| 7/10/2002 | 1.0178 | |
| 7/12/2002 | | 1.852 |
| 7/24/2002 | 10.057 | |
| 7/31/2002 | 6.4011 | |
| 8/1/2002 | 23.688 | |
| 8/6/2002 | 22.488 | |
| 8/12/2002 | | 10.219 |
| 8/25/2002 | 39.705 | |
| 8/31/2002 | 37.406 | |
| 9/1/2002 | 71.573 | |
| 9/4/2002 | 71.093 | |
| 9/7/2002 | | -1.0777 |
| 9/23/2002 | 88.124 | |
| 9/27/2002 | | 7.0019 |
| 10/7/2002 | 37.223 | |
| 10/8/2002 | | 4.77 |
| 10/20/2002 | 44.13 | |
| 10/21/2002 | | 7.4265 |
| 10/31/2002 | 34.479 | |
| 11/1/2002 | 72.569 | |
| 11/8/2002 | | -1.8044 |
| 11/9/2002 | 66.836 | |
| 11/14/2002 | 67.112 | |
| 11/26/2002 | | 9.1322 |
| 12/7/2002 | 14.971 | |
| 12/11/2002 | | 4.0458 |
| 12/13/2002 | | 4.0743 |
| 12/16/2002 | 16.74 | |
| 12/31/2002 | -0.31832 | |
| 1/1/2003 | 1.7767 | |
| 1/8/2003 | | -1.8829 |
| 1/31/2003 | | 9.9921 |
| 2/1/2003 | | 9.9666 |
| 2/7/2003 | | 10.49 |
| 2/24/2003 | | -2.3284 |
| 2/28/2003 | -72.375 | |
| 3/1/2003 | -2.4867 | |
| 3/4/2003 | -3.0065 | |
| 3/18/2003 | 5.5706 | |
| 3/29/2003 | | 7.4058 |
| 4/2/2003 | -4.3533 | |
| 4/17/2003 | 4.2222 | |
| 5/2/2003 | -5.6178 | |
| 5/7/2003 | | -0.026177 |
| 5/16/2003 | 3.2456 | |
| 5/28/2003 | | 6.0334 |
| 6/1/2003 | -29.59 | |
| 6/14/2003 | | -1.9162 |
| 6/18/2003 | -10.574 | |
| 6/20/2003 | | -0.92177 |
| 6/25/2003 | -11.768 | |
| 7/2/2003 | | -2.6374 |
| 7/14/2003 | 3.8503 | |
| 7/24/2003 | | 5.5069 |
| 7/29/2003 | -5.1896 | |
| 8/15/2003 | 30.171 | |
| 8/22/2003 | | -4.309 |
| 8/25/2003 | 27.703 | |
| 9/12/2003 | 43.109 | 5.3503 |
| 9/25/2003 | 37.016 | |
| 9/28/2003 | | 2.886 |
| 10/4/2003 | | 3.1032 |
| 10/10/2003 | 46.721 | |
| 10/20/2003 | | -1.7628 |
| 10/27/2003 | 35.224 | |
| 11/5/2003 | 77.888 | 0.0047447 |
| 11/6/2003 | | -0.07118 |
| 11/25/2003 | | 4.3105 |
| 12/1/2003 | 23.941 | |
| 12/5/2003 | 24.913 | |
| 12/23/2003 | | -2.9934 |
| 12/27/2003 | 3.3766 | |
| 1/2/2004 | 4.4056 | |
| 1/18/2004 | | 8.3826 |
| 1/26/2004 | -17.72 | |
| 1/31/2004 | -16.658 | |
| 2/13/2004 | | -7.9399 |
| 2/27/2004 | | 4.2253 |
| 2/28/2004 | -68.972 | |
| 3/4/2004 | -32.08 | 2.4746 |
| 3/17/2004 | | 5.7736 |
| 3/22/2004 | -46.126 | |
| 4/5/2004 | -38.826 | |
| 4/13/2004 | | -0.57219 |
| 4/19/2004 | -47.492 | |
| 4/23/2004 | | 5.2319 |
| 5/5/2004 | | -0.81594 |
| 5/6/2004 | -34.703 | |
| 5/14/2004 | | 1.4443 |
| 5/17/2004 | -38.886 | |
| 6/6/2004 | -19.917 | |
| 6/12/2004 | | -3.5549 |
| 6/13/2004 | -21.549 | |
| 7/7/2004 | 2.079 | 7.5255 |
| 7/12/2004 | 1.4422 | |
| 8/2/2004 | | -1.2131 |
| 9/4/2004 | 78.962 | |
| 9/11/2004 | 78.319 | |
| 9/14/2004 | | 9.8085 |
| 9/30/2004 | 94.554 | |
| 10/2/2004 | | -2.1026 |
| 10/12/2004 | | 0.63555 |
| 10/15/2004 | 85.169 | |
| 10/19/2004 | | -0.19246 |
| 10/26/2004 | 89.565 | |
| 11/12/2004 | | 5.4216 |
| 12/1/2004 | | -2.3638 |
| 12/28/2004 | | 11.098 |
| 1/14/2005 | | 1.6975 |
| 1/15/2005 | | 1.7456 |
| 1/16/2005 | | 1.728 |
| 1/23/2005 | | 1.9433 |
| 2/5/2005 | | 0.58157 |
| 2/28/2005 | -105.64 | |
| 3/1/2005 | -0.84346 | |
| 3/10/2005 | | 12.574 |
| 3/11/2005 | -7.8665 | |
| 3/26/2005 | 1.5037 | |
| 3/31/2005 | | -4.16 |
| 4/9/2005 | -50.484 | |
| 4/16/2005 | | 6.8204 |
| 4/25/2005 | -38.451 | |
| 4/28/2005 | | 2.803 |
| 5/7/2005 | -43.291 | |
| 5/12/2005 | | 6.0747 |
| 5/26/2005 | -25.406 | |
| 6/1/2005 | | 1.0214 |
| 6/3/2005 | | 1.0962 |
| 6/4/2005 | -27.622 | 1.0687 |
| 6/25/2005 | -5.1874 | |
| 6/28/2005 | | 7.8053 |
| 7/2/2005 | -6.1702 | |
| 7/24/2005 | | 0.77917 |
| 7/25/2005 | 17.632 | |
| 8/1/2005 | 16.508 | |
| 8/3/2005 | | 4.0725 |
| 8/14/2005 | | 0.82293 |
| 8/22/2005 | 37.621 | |
| 8/31/2005 | 34.879 | |
| 9/2/2005 | | 6.7547 |
| 9/20/2005 | 91.059 | |
| 9/24/2005 | | -3.234 |
| 10/3/2005 | 42.925 | |
| 10/17/2005 | 51.375 | |
| 10/20/2005 | | 4.1569 |
| 10/21/2005 | | 4.0999 |
| 10/31/2005 | 38.853 | |
| 11/1/2005 | 77.453 | |
| 11/6/2005 | 75.231 | |
| 11/9/2005 | | 6.3224 |
| 11/11/2005 | 75.691 | |
| 12/7/2005 | | 3.7075 |
| 12/8/2005 | | 3.7511 |
| 12/9/2005 | | 3.7398 |
| 12/23/2005 | | 9.1266 |
| 1/15/2006 | | -1.7744 |
| 2/16/2006 | | 5.0022 |
| 2/17/2006 | | 4.9667 |
| 2/18/2006 | | 4.9697 |
| 2/26/2006 | | 4.3054 |
| 2/28/2006 | -76.981 | |
| 3/5/2006 | | 5.189 |
| 3/15/2006 | 4.0413 | |
| 3/22/2006 | | -5.164 |
| 3/29/2006 | -4.154 | |
| 3/31/2006 | -3.7555 | |
| 4/1/2006 | -46.239 | |
| 4/14/2006 | -36.14 | |
| 4/15/2006 | | 3.7998 |
| 4/27/2006 | -41.946 | |
| 5/7/2006 | | -1.9257 |
| 5/13/2006 | 6.9536 | |
| 5/26/2006 | | 3.5458 |
| 5/27/2006 | -2.421 | |
| 5/31/2006 | -1.2046 | |
| 6/1/2006 | -24.076 | |
| 6/6/2006 | | 1.6216 |
| 6/14/2006 | -9.7434 | |
| 6/23/2006 | -11.766 | 6.547 |
| 7/11/2006 | 4.6868 | |
| 7/16/2006 | | 0.79062 |
| 7/26/2006 | -5.3102 | |
| 8/3/2006 | | 5.6393 |
| 8/12/2006 | 27.586 | |
| 8/16/2006 | | 1.4759 |
| 8/21/2006 | 24.614 | |
| 8/25/2006 | | 2.9892 |
| 9/11/2006 | 79.241 | 0.72211 |
| 9/12/2006 | | 0.73932 |
| 9/13/2006 | | 0.6676 |
| 9/20/2006 | 76.654 | |
| 9/30/2006 | 83.882 | |
| 10/1/2006 | 42.322 | |
| 10/7/2006 | 46.464 | |
| 10/18/2006 | | 5.7386 |
| 10/23/2006 | 36.176 | |
| 10/29/2006 | | 5.3427 |
| 11/3/2006 | 80.444 | 5.3607 |
| 11/27/2006 | | -0.33099 |
| 12/1/2006 | 27.545 | |
| 12/2/2006 | 27.722 | |
| 12/13/2006 | | 7.965 |
| 12/24/2006 | 5.7105 | |
| 12/30/2006 | 7.0053 | |